\documentclass[12pt]{article}
\usepackage{amsmath,amssymb,amsthm,latexsym,graphicx}
\newtheorem{theorem}{Theorem}
\newtheorem{lemma}[theorem]{Lemma}

\newtheorem{definition}[theorem]{Definition}

\newcommand{\CC}{\mathbb{C}}
\newcommand{\RR}{\mathbb{R}}
\newcommand{\ZZ}{\mathbb{Z}}
\newcommand{\TT}{\mathbb{T}}

\newcommand{\GG}{\Gamma}
\newcommand{\ee}{\mathbf{e}}

\begin{document}

\title{On the structure of eigenfunctions corresponding to embedded eigenvalues\\
 of locally perturbed periodic graph operators}
\author{Peter Kuchment\\
Department of Mathematics,  Texas A\&M University\\
 College Station, TX,  77843-3368\\
 kuchment@math.tamu.edu\\
 Boris Vainberg\\
 Mathematics Department, University of North Carolina\\
 Charlotte, NC 28223\\
brvainbe@email.uncc.edu}
 \date{}
\maketitle

\begin{abstract}
The article is devoted to the following question. Consider a
periodic self-adjoint difference (differential) operator on a
graph (quantum graph) $G$ with a co-compact free action of the
integer lattice $\ZZ^n$. It is known that a local perturbation of
the operator might embed an eigenvalue into the continuous
spectrum (a feature uncommon for periodic elliptic operators of
second order). In all known constructions of such examples, the
corresponding eigenfunction is compactly supported. One wonders
whether this must always be the case. The paper answers this
question affirmatively. What is more surprising, one can estimate
that the eigenmode must be localized not far away from the
perturbation (in a neighborhood of the perturbation's support, the
width of the neighborhood determined by the unperturbed operator
only).

The validity of this result requires the condition of
irreducibility of the Fermi (Floquet) surface of the periodic
operator, which is expected to be satisfied for instance for
periodic Schr\"{o}dinger operators.

{\bf 2000 Mathematics Subject Classification:} 05C90, 34B45,
34L05, 34L40, 35P05, 47A10, 58J50.

{\bf Keywords:} graph, quantum graph, spectrum, embedded
eigenvalue, Fermi surface, Floquet theory

\end{abstract}

\section{Introduction}

Difference equations on graphs and differential equations on
quantum graphs, even when they resemble Laplace or Schr\"{o}dinger
operators in many regards, lack one important property of second
order elliptic operators, namely uniqueness of continuation.
Uniqueness of continuation states that any solution of an elliptic
second order equation $Au=0$ that vanishes on an open set, is
identically zero. This is known to be an extremely important
property with many implications, in particular in spectral theory.
It is also known that elliptic equations of higher orders do not
necessarily possess such a property \cite{Plis}, which leads to
some weird spectral examples as well (e.g., \cite{Kuch_book}).

This absence of uniqueness of continuation leads for instance to
the following possibilities: a periodic ``elliptic second order''
operator on a graph (quantum graph) with a co-compact action of an
abelian group can have non-empty pure point spectrum (bound
states) \cite{KuchDifference}; this is an absolute no-no in the
continuous case, see \cite{Kuch_book,RS,Thomas} and references
therein. It is easy to explain this effect for instance as
follows: assuming that one has a compact graph with an
eigenfunction of the discrete Laplacian that has a zero at some
vertex, one can attach this graph by that vertex to any other
graph and extend the function as zero still keeping it an
eigenfunction. This attachment can also be done in a periodic
manner. Such constructions yield these ``strange'' eigenfunctions
generated by compactly supported ones. Indeed, it has been shown
that all such bound states on periodic graphs are in fact
generated by the compactly supported ones
\cite{KuchDifference,Kuch_graphs2}. It is interesting to note that
Laplace operator on the Cayley graph of an infinite discrete group
can even have solely pure point spectrum \cite{dicks,GrigZuk}.

Using the described above attachment procedure, one can also
easily construct examples when a localized perturbation of a
periodic operator does embed an eigenvalue into absolutely
continuous spectrum, which is also expected to be impossible in
the continuous situation (albeit this is completely proven in
dimensions one only \cite{Rofe1,Rofe2} with just a single result
in higher dimension available \cite{KuchVain1,KuchVain2}). The aim
of this paper is to see what can be said about the eigenfunctions
corresponding to such embedded eigenvalues. We show not only that
such an eigenfunction must be compactly supported, but that its
support must be contained in a finite width neighborhood of the
support of the perturbation, the width dependent on the
unperturbed operator only. Thus, effect of the perturbation seems
to be of an extremely short range, when being on the absolutely
continuous spectrum of the periodic background.

In the next section we introduce the necessary notions and state
and prove the main result for the case of periodic combinatorial
graphs (Theorem \ref{T:main}). The following section contains
formulation and the proof of the analogous result for the quantum
graph case (Theorems \ref{T:main_quantum} -
\ref{T:main_quantum2}). The paper ends with a brief section
containing some final remarks.

\section{Combinatorial graph case}

Consider an infinite combinatorial graph $\GG$ with the set of
vertices $V$ and a faithful co-compact action of the free abelian
group $G=\ZZ^n$ (i.e., the space of $G$-orbits is a finite graph).
In fact, in this section we can think of $\GG$ just as of a
discrete set $V$ of vertices; the graph structure is not truly
relevant here, albeit the main operators of interest usually come
from graphs (e.g., graph Laplacian \cite{Chung}). Without loss of
generality, the reader may think of the graph as a discrete subset
of $\RR^n$ invariant with respect to all integer shifts. We also
consider a $G$-periodic finite difference operator $A$ of a finite
order acting on $l_2(V)$. Here $l_2(V)$ is the space of all square
summable functions on $\GG$ (i.e., on $V$). The words ``finite
difference operator of a finite order'' mean that the value of
$Au$ at any vertex $v$ involves the values of $u$ at a finitely
many other vertices (due to periodicity, the number of these
vertices is uniformly bounded). Such periodic operators are
clearly bounded in $l_2(V)$.

We will fix a (finite) fundamental domain $W$ for the action of
$G=\ZZ^n$ on $V$.

Consider for instance the graph below that is a $\ZZ^2$-periodic
sub-graph of $\RR^2$, with the fundamental domain $W$ indicated.
\begin{figure}[ht]\label{F:graph}
\begin{center}\scalebox{0.7}{\includegraphics{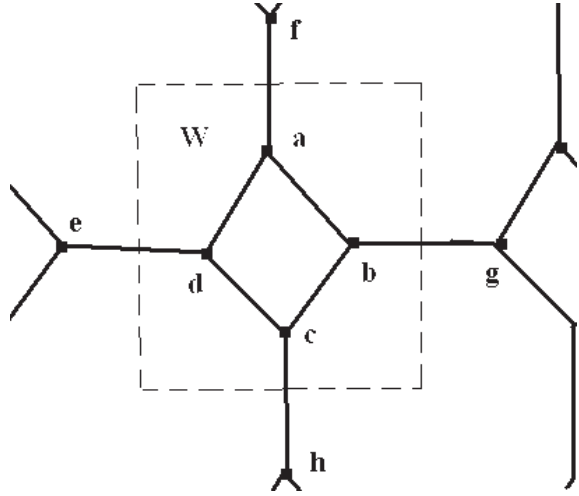}}
\caption{A periodic graph}\end{center}
\end{figure}
An example of a periodic difference operator here could be the
version of the Laplace operator that subtracts from the value of a
function at a vertex its average value at all vertices adjacent to
this one: $\Delta f(v)=f(v)-\frac{1}{d_v}\sum\limits_{u \sim
v}f(u)$, where $d_v$ is the degree of the vertex $v$. This
operator is clearly a finite difference operator periodic with
respect to the group action.

We will need to measure the sizes of finite subsets $S\subset \GG$
by the number and locations of the shifts of the fundamental
domain $W$ that are needed to cover $S$. Given a finite subset $S$
of $\GG$ we will call its {\em radius} the number
\begin{equation}\label{E:radius}
r(S)=\min \{N\in \ZZ^+\,|\, S\subset
\mathop{\cup}\limits_{\gamma\in [-N,N]^n\subset \ZZ^n} \gamma W\}.
\end{equation}

We will also need to define two notions of support of a finite
difference periodic operator $A$. First of all, let $x$ be a
vertex of $\GG$. Then we introduce the notion of the $x$-support
of $A$ as follows
\begin{equation}\label{E:supp_operator}
    supp_x(A)=\{v\in V\,|\, (A\delta_v)(x)\neq 0\}.
\end{equation}
Here $\delta_v$ is the delta function on $V$ supported at the
vertex $v$, i.e. $\delta_v(u)=\delta_{u,v}$ for $u,v\in V$. To put
it differently, the $x$-support of $A$ consists of all points $v$,
values at which of a function $\psi$ influence the values of
$A\psi$ at $x$. We also define the $W$-support of $A$ as
\begin{equation}\label{E:x-supp_operator}
    supp_W(A)=\mathop{\cup}\limits_{x\in W}supp_x(A)\\
    =\{v\in V|\,\,\, A\delta_v |_W\neq 0\}.
\end{equation}
In other words, the $W$-support of $A$ consists of all points $v$
values at which of a function $\psi$ influence the values of
$A\psi$ on $W$.

As always, dealing with a periodic problem, it is advantageous to
use the basic transforms of  {\em Floquet theory} (e.g.,
\cite{Kuch_book,RS}). Namely, for any compactly supported (or
sufficiently fast decaying) function $f(v)$ on $V$, we define its
{\em Floquet transform} as follows:
\begin{equation}\label{E:Floquet}
  f(v)\mapsto \hat{f}(v,z)=\sum\limits_{g\in\ZZ^n} f(gv)z^{-g},
\end{equation}
where $gv$ denotes the action of $g\in\ZZ^n$ on the point $v\in
v$, $z=(z_1,...,z_n)\in (\CC^*)^n=(\CC\backslash 0)^n$, and
$z^g=z_1^{g_1}\times...\times z_n^{g_n}$. This is clearly just the
Fourier transform on the group $G$ of periods.

One can notice the easily verifiable cyclic (or Floquet) identity
\begin{equation}\label{E:cyclic}
  \hat{f}(gv,z)=z^{g}\hat{f}(v,z)
\end{equation}
satisfied for any $v\in V$ and $g \in G$. The vector $z$ is
sometimes called {\em Floquet multiplier}, and if being
represented as $z=e^{ik}=(e^{ik_1},e^{ik_2},...,e^{ik_n})$, vector
$k$ is said to be the {\em quasi-momentum} (e.g.,
\cite{AM,Kuch_book,RS}).

Relation (\ref{E:cyclic}) implies that in order to know all the
values of the function $\hat{f}(v,z)$, it is sufficient to know
them for only one representative $v$ from each $G$-orbit, i.e. for
$v$ from a fundamental domain of the $G$-action\footnote{In some
cases one has to take a more sophisticated approach and treat
$\hat{f}(v,z)$ as a section of a naturally defined (depending on
$z$) line bundle over $\GG /G$.}. Thus, we fix such a fundamental
domain $W$ (which is a finite set (graph)) and consider only $v\in
W$ in $\hat{f}(v,z)$. We will also denote $\hat{f}(v,z)$ by
$\hat{f}(z)$, where the latter expression is a function on $W$
depending on the parameter $z$.

The following statement is immediate:
\begin{lemma}\label{L:Laurent} The images under the Floquet transform
of the compactly supported functions on $\GG$ are exactly all
finite Laurent series\footnote{By {\em Laurent series} we mean
here expansions into powers $z^g$ with $g\in G =
 \ZZ^n$.} in $z$ with coefficients in $\CC^{|W|}$. Moreover, for a
 compactly supported function $f$,
 the Laurent series of $\widehat{f}$ includes only powers $z^g$ that satisfy
 \begin{equation}\label{E:powers}
    \|g\|_\infty:=\max (|g_j|)\leq r( supp(f)),
\end{equation}
where $r(S)$ is the radius of a set $S$ introduced in
(\ref{E:radius}).
\end{lemma}

We will also need the unit torus
$$
\TT^n=\{z\in\CC^n\,|\,|z_j|=1,j=1,...,n\}\subset\CC^n.
$$
It is well known and easy to prove \cite{Eastham,Kuch_book,RS}
that the Floquet transform (\ref{E:Floquet}) extends to an
isometry (up to a possible constant factor) between $l_2(V)$ and
$L_2(\TT^n,\CC^{|W|})$.

After the Floquet transform, $A$ becomes the operator of
multiplication in $L_2(\TT^n,\CC^{|W|})$ by a rational $|W|\times
|W|$ matrix function $A(z)$. To make this clearer, let us compute
for the graph shown in Fig. 1 and the Laplace operator $\Delta$
its value $\Delta u$ on a function $u$ that satisfies
(\ref{E:cyclic}). We notice that for such a function one has
$u(f)=z_2u(c), u(g)=z_1u(d), u(h)=z_2^{-1}u(a),
u(e)=z_1^{-1}u(b)$. Thus, writing the values of $u$ as a vector
$(u(a),u(b),u(c),u(d))$, the action of $\Delta$ on $u|_W$ becomes
multiplication by the matrix $A(z)$
\begin{equation}\label{E:matrix}
   \left(%
\begin{array}{cccc}
  1 & -1/3 & -1/3z_2 & -1/3 \\
 -1/3 & 1 & -1/3 & -1/3 z_1\\
  -1/3z_2^{-1} & -1/3 & 1 & -1/3 \\
  -1/3 & -1/3z_1^{-1} & -1/3 & 1 \\
\end{array}%
\right)
\end{equation}

To formulate our result, we need to introduce another notion.

\begin{definition}\label{D:Floquet_surf}
{\em Let $\lambda \in \CC$. We call} the  Floquet surface
$\Phi_{A,\lambda}\subset (\CC^*)^n$ of the operator $A$ at the
energy $\lambda$ {\em the set of all $z\in (\CC^*)^n$ such that
the matrix $A(z)-\lambda$ is not invertible (i.e., $\det
(A(z)-\lambda)=0$).}
\end{definition}

The term {\em Floquet surface} is non-standard. If one considers
quasi-momenta $k$ instead of the Floquet multipliers $z$, one
arrives to the standard in solid state physics and theory of
periodic equations notion of {\em Fermi surface} $F_{A,\lambda}$
\cite{AM,Kuch_book}. So, the Floquet surface is the Fermi one with
the natural periodicity with respect to quasimomenta $k$ being
factored out.

It is clear from the definition that the Floquet surface is an
algebraic subset of dimension $n-1$ in $\CC^n$. We also look at
its intersection with the torus $\TT^n$
$$
\Phi^R_{A,\lambda}=\Phi_{A,\lambda} \cap \TT^n,
$$
which we will call the {\em real Floquet surface}. The name comes
from the fact that its correspond to real quasimomenta from the
Fermi surface.

The following standard fact \cite{Eastham,Kuch_book,RS} is easy to
prove:
\begin{lemma}\label{L:spectr}
The point $\lambda$ belongs to the spectrum of the operator $A$ if
and only if $\Phi^R_{A,\lambda}\neq \emptyset$.
\end{lemma}

We will also need to introduce some additional notions originating
from the solid state physics \cite{AM}. Consider for any
$z\in\TT^n$ the operator $A_z$ which is the restriction of $A$ to
the space of all (not square integrable) functions $f$ satisfying
the cyclic condition (\ref{E:cyclic}). This is a finite
dimensional self-adjoint operator that has a finite spectrum
$\{\lambda_j(z)\}$, which can be considered as the graph of a
multiple-valued function $\sigma(A_z)$. This function is said to
be the {\em dispersion relation} and its graph the {\em dispersion
curve}. The preceding Lemma says that the range of this function
coincides with the spectrum of $A$ in $l_2(\GG)$. Arranging the
eigenvalues in non-decreasing order splits this curve into
continuous (in fact, piecewise-analytic
\cite{Kuch_book,RS,Wilcox}) branches $\lambda_j(z)$. Their ranges
are finite closed segments of the spectral axis called {\em
spectral bands}, union of which comprise the whole spectrum
$\sigma(A)$. This is the so called band-gap structure of the
spectrum \cite{Eastham,Kuch_book,RS}.

The (complex) Floquet surface is never empty, but when $\lambda$
changes, it moves around. The lemma says that whenever it hits the
torus $\TT^n$, $\lambda$ belongs to the spectrum. It is natural to
expect that when $\lambda$ is a generic point in the interior of
the spectrum, then the real Floquet surface will be a variety of
the maximal possible real dimension $n-1$ in the torus. This is
confirmed by the following statement.

\begin{lemma}\label{L:inside_band}
If $\lambda$ belongs to the interior of a spectral band of the
operator $A$, then the real Floquet surface has a part which is a
smooth $n-1$-dimensional hyper-surface in $\TT^n$.
\end{lemma}
{\bf The sketch of the proof of the Lemma} (see more details in
\cite{KuchVain1}). Let $\lambda$ belongs to the interior of the
band formed by the branch $\lambda_j(z)$. Then function
$\lambda_j(z)-\lambda$ changes sign on $\TT^n$. Thus, the real
Floquet surface separates $\TT^n$. Now the known analytic
structure of the Floquet surface (it is an analytic, or even
algebraic in discrete situation set) leads to the conclusion of
the lemma (see more details of this part of the argument in
\cite{KuchVain1}).\qed

In what follows, we will need to know that the Floquet surface of
$A$ is irreducible as an algebraic variety\footnote{We remind the
reader what this means: the variety cannot be represented as the
union of two strictly smaller algebraic varieties.}. This is not
always true, but for instance it is conjectured that this is
always true if $A$ is the discrete Schr\"{o}dinger operator on
$\ZZ^2$ with potential periodic with respect to a sublattice
\cite{Gieseker}. This probably is true in any dimension. It was
shown in \cite{Gieseker} that in $2D$ irreducibility holds for all
but a finitely many of values of the spectral parameter $\lambda$.
Examples of some separable cases when irreducibility has been
proven can be found in \cite{BKT,Gieseker,KuchVain1,KuchVain2}.

After all this preparation, let us now move to the formulation of
the main problem we address in this paper. Consider any {\em
local} difference operator $B$, i.e. such that its action on a
function $u$ involves only the function's values on a finite set
$S\subset \GG$ and the resulting function $Bu$ is supported on $S$
as well. We are interested in the perturbation of the spectrum of
$A$ that occurs when the operator is perturbed by adding $B$:
$A+B$. If we assume at this point that $A$ is self-adjoint, it is
a general fact that an additional point spectrum might arise
(e.g., \cite{RS}). It is also the common expectation that this
impurity spectrum due to local perturbations should not be
embedded into the continuous spectrum of $A$, which is proven for
perturbations of a homogeneous background (see the book
\cite{Eastham2} for a detailed survey). In the case of localized
perturbations of a periodic background, absence of embedded
eigenvalues is proven for periodic Schr\"{o}dinger operators in
$1D$ \cite{Rofe1,Rofe2}. Albeit the same must surely be true in
any dimension, the problem in dimensions higher than $1$ is hard
and only one limited result is known \cite{KuchVain1,KuchVain2}.
In the discrete (graph) situation, embedded eigenvalues can arise
very easily, due to non-trivial graph topology. Examples of such
compactly supported eigenfunctions can be easily constructed using
the attachment procedure described before. One might want to ask
whether compactness of support of the eigenfunctions corresponding
to embedded eigenvalues is the only possibility, and if yes,
whether there are any {\em a priori} bounds on the size of their
supports. Somewhat surprising answer is given by the following
result.
 \begin{theorem}\label{T:main}
Let $B$ be a local perturbation supported on a finite set
$S\subset \GG$ (i.e., $supp (Bf)\subset S$ for any $f$) of a
periodic operator $A$. Let $\lambda$ belong to the interior of a
spectral band of $A$, the corresponding Floquet surface be
irreducible, and $\lambda$ be an embedded eigenvalue for $A+B$.
Then the corresponding eigenfunction $f\in l_2(V)$ of $A+B$ is
compactly supported and moreover,
$$
r(supp\,f) < r(S)+r(supp_W(A))(2|W|+1)).
$$
The constant $r(supp_W(A))(2|W|+1))$ can be usually improved for
specific periodic difference operators $A$. (Here $supp_W(A)$ is
defined in (\ref{E:supp_operator}).)
\end{theorem}
So, the effect of the impurity seems to be of very short range.
This theorem will be deduced from the following more general
statement:
\begin{theorem}\label{T:nonhomogeneous}
Let $\lambda$ belong to the interior of a spectral band of $A$,
the corresponding Floquet surface be irreducible, and $\psi$ be a
compactly supported function on the graph. Assume that the
equation $Au-\lambda u=\psi$ has an $l^2$-solution $u$. Then $u\in
l_2(V)$ is compactly supported and moreover,
$$
r(supp\,f) < r(supp\,\psi)+r(supp_W(A))(2|W|+1)),
$$
where the constant $r(supp_W(A))(2|W|+1))$ can be usually improved
for specific periodic difference operators $A$.
\end{theorem}

{\bf Proof of Theorem \ref{T:nonhomogeneous}}. Since function
$\psi$ is compactly supported, its Floquet transform
$\widehat{\psi}(v,z)=\sum \psi_g z^g$ is a Laurent polynomial with
degrees $g$ bounded by $\|g\|_\infty:=\max\limits_i |g_i|\leq
r:=r(supp(\psi))$. Let us denote by $\ee$ the vector
$(1,...,1)\in\ZZ^n$ and introduce $R:=r(supp_W(A))$. We can
represent $\widehat{\psi}(z)$ as $z^{-r\ee}\phi(z)$, where
$\phi(z)=z^{r\ee}\widehat{\psi}(z)$ is a polynomial that involves
only (non-negative) degrees $g\in\ZZ^n$ with $\|g\|_\infty \leq
2r$.

Taking Floquet transform in the equation (\ref{E:inhom}), we
rewrite it as
\begin{equation}\label{E:Fl_inhom}
(A(z)-\lambda)\widehat{f}(z)=z^{-r\ee}\phi(z).
\end{equation}
We can rewrite the Laurent matrix function $A(z)-\lambda$ as
$z^{-R\ee} A_1(z,\lambda)$, where the matrix function
$A_1(z,\lambda)$ is a polynomial in $z$ involving only the powers
$z^g$ with $\|g\|_\infty\leq 2R$. Then its inverse can be
represented as $z^{R\ee}\frac{B(z)}{\Delta(z)}$, where $B(z)$ is a
matrix polynomial (transposed to the co-factor matrix of $A_1$)
and $\Delta(z)$ is a scalar polynomial (determinant of $A_1$),
which vanishes exactly on the Floquet surface. Thus,
\begin{equation}\label{E:divide}
    \widehat{f}(z)=z^{(R-r)\ee}\frac{B(z)\phi(z)}{\Delta(z)}.
\end{equation}
Notice that $B$ involves only powers $z^g$ with $\|g\|_\infty\leq
2R(|W|-1)$. We know that $\widehat{f}(z)$ is an $L^2$-function on
$\TT^n$. On the other hand, the right hand side of
(\ref{E:divide}) is, up to the factor $z^{(R-r)\ee}$, the ratio of
two holomorphic polynomials in $\CC^n$. We also know that zeros of
the denominator $\Delta(z)$ in $(\CC^*)^n$are irreducible and
intersect the torus $\TT^n$ over an $(n-1)$-dimensional variety.
This means that unless the numerator $B\phi$ vanishes on $\TT^n$
at these zeros to their degrees, the ratio has a singularity that
is not square integrable. Thus, the numerator vanishes to that
degree, and due to the irreducibility of zeros, the same is true
for all zeros in $(\CC^*)^n$ (see \cite{KuchVain1} for the details
of this simple argument). If there were no zeros of the
denominator in $\CC^n\backslash (\CC^*)^n$, then, as a corollary
of Hilbert's Nullstellensatz, the ratio would be a holomorphic
polynomial of $z$. We cannot, however, exclude existence of a
factor like $z^q,\,q\in(\ZZ^+)^n$ in $\Delta(z)$. If it does
exist, we have $\|q\|_\infty \leq 2R|W|$ (since each term in
$\Delta$ is like that). Factoring this power out, we represent
$\Delta$ as $z^q \Delta_1(z)$, where zeros of $\Delta$ and
$\Delta_1$ in $(\CC^*)^n$ are the same (including their orders),
and thus our ratio is a holomorphic polynomial times $z^{-q}$.
Notice that division does not increase the degree of a polynomial
with respect to any variable. The degree of $\phi(z)$ has been
estimated as $\|g\|_\infty \leq 2r$. The additional degree
acquired during multiplication by $B$ and division by $\Delta_1$
does not exceed $2R(|W|-1)$. Thus, the ratio $B\phi/\Delta_1$ is a
polynomial involving the degrees $z^g$ with $\|g\|_\infty \leq
2r+2R(|W|-1)$ only. One can calculate now that the effect of the
outside factor of $z^{(R-r)\ee}$ and of $z^{-q}$ coming from the
denominator is to reduce the expression to a Laurent polynomial
with degrees $z^g,\,g\in\ZZ^n$ involved with $\|g\|_\infty \leq
r+R(2|W|+1)$. We see that $\widehat{f}(z)$ is a Laurent polynomial
which contains powers of $z$ estimated by
$r(\psi)+r(supp_W(A))(2|W|+1)$. Reversing the Floquet transform,
we get the statement of the theorem. \qed

Let us now address the {\bf proof of Theorem \ref{T:main}}, which
is rather simple. Indeed, the conditions of the theorem imply the
equality $Af+Bf=\lambda f$, or in a form more convenient for us,
\begin{equation}\label{E:inhom}
(A-\lambda)f=-Bf:=\psi.
\end{equation}
The function $\psi (v)$ is supported on $S$. Thus Theorem
\ref{T:nonhomogeneous} applies and proves the statement. \qed

\section{Quantum graph case}
We now switch to the case of a perturbed periodic quantum graph.
We will remind the reader the main definitions concerning quantum
graphs. One can find more details concerning quantum graphs in
\cite{KoS1}, \cite{Kush}-\cite{Kuch_graphs2}, \cite{QGraphs}. A
quantum graph $\GG$ has each its edge $e$ equipped with a
coordinate $x_e$ (when no confusion is possible, we use just $x$
instead). This coordinate identifies $e$ with a segment $[0,l_e]$
of the real line. We will also assume that a Schr\"{o}dinger
operator $H=-\frac{d^2}{dx^2}+V(x), V\in L^2_{loc}(\GG)$ with
appropriate vertex conditions (all such self-adjoint conditions
are described in \cite{Harmer,KoS1,Kuch_graphs1}) is defined on
$\GG$. The results of this section hold for any of such
conditions, however just for simplicity of presentation we will
assume that the conditions at each vertex are the ``standard''
Neumann-Kirchhoff ones:
\begin{equation}\label{E:conditions}
    f \mbox{ is continuous and } \sum\frac{df}{dx_e}=0 \mbox{ at each
    vertex},
\end{equation}
where the sum is taken over all edges incident with the vertex and
the derivatives are taken away from the vertex.

As in the previous section, we assume that the graph is acted upon
freely and co-compactly by the group $G=\ZZ^n$ that leaves the
graph structure (including edges' lengths) and the Hamiltonian $H$
invariant. We use the same letter $W$ as before for a fundamental
domain of this action.

One can now introduce the notions of the Floquet transform and
Floquet variety of $H$ analogously to the way it was done in the
previous section. For instance,
\begin{definition}\label{D:Floquet_quantum}
{\em Let $\lambda \in \CC$. We call} the Floquet surface
$\Phi_{H,\lambda}\subset (\CC^*)^n$ of the operator $H$ at the
energy $\lambda$ {\em the set of all $z\in (\CC^*)^n$ such that
the equation $(H-\lambda)u=0$ has a non-trivial solution $u$ that
is cyclic with the Floquet multiplier $z$, i.e. such that
$u(gx)=z^gu(x)$, where $x\in \GG$ and $g\in \ZZ^n$. Here, as
before, $\CC^*=\CC \backslash \{0\}$.}
\end{definition}
We can also introduce the same assumption about irreducibility.

The main result of this section is the quantum graph analog of
Theorem \ref{T:main}.
\begin{theorem}\label{T:main_quantum}
Let $w(x)\in L^2(\GG)$ be supported on a finite set $S$ of edges.
Let $\lambda$ belong to the interior of a spectral band of $H$,
the corresponding Floquet surface be irreducible, and $\lambda$ be
an embedded eigenvalue for $H+w$. Then the corresponding
eigenfunction $f\in L_2(\GG)$ of $H+w$ is compactly supported and
moreover,
$$
r(supp\,f) < r(\widetilde{S})+C.
$$
Here $C$ is a constant depending on the unperturbed operator $H$
only and for any set of vertices $S$  we define $\widetilde{S}$ as
the set of all vertices that are either in $S$, or adjacent to the
ones in $S$.
\end{theorem}

It is possible to give some explicit estimates for the constant
$C$, similarly to how it was done for the discrete case. Here,
however, the situation depends on whether or not $\lambda$ is the
Dirichlet eigenvalue of $H$ on an edge of the graph.

\begin{theorem}\label{T:main_quantum1}
Let $w(x)\in L^2(\GG)$ be supported on a finite set $S$ of edges.
Assume that $\lambda$ belongs to the interior of a spectral band
of $H$ and is not a Dirichlet eigenvalue on any of the edges, the
corresponding Floquet surface is irreducible, and $\lambda$ is an
embedded eigenvalue for $H+w$. Then the corresponding
eigenfunction $f\in L_2(\GG)$ of $H+w$ is compactly supported and
moreover,
$$
r(supp\,f) <
r(\widetilde{S})+r(\widetilde{W})(2|\widetilde{W}|+1).
$$
Here we consider $W$ as a set of vertices.
\end{theorem}

In case when $\lambda$ does belong to the Dirichlet spectrum of at
least one of the edges, the situation is different, and one needs
to modify the graph somewhat. We would like to guarantee that
$\lambda$ does not contain any points of Dirichlet spectra of $H$
and of $H+w$ on any of the edges of the graph. This is easy to
achieve by introducing ``fake'' vertices. Indeed, if all the edges
are sufficiently short, this condition is satisfied. Now, modulo
the periodicity, there are only finitely many edges in the graph.
Hence, one can introduce a finite set of periodic families of
interior points on the edges, such that including these points as
new vertices of degree two, one achieves the desired result.
Imposing Neumann-Kirchhoff conditions (\ref{E:conditions}) at
these vertices (in the case of a vertex of degree two they just
mean continuity of the function and its derivative), one makes
sure that they do not influence the spectra of $H$ and of $H+w$ at
all. This reduces the situation to the case of Theorem
\ref{T:main_quantum1}, however with an increased number of
vertices in the fundamental domain. Let us call this new set
$W_1$. Then clearly Theorem \ref{T:main_quantum1}, if proven,
implies the next theorem, and thus also Theorem
\ref{T:main_quantum}:

\begin{theorem}\label{T:main_quantum2}
Let $w(x)\in L^2(\GG)$ be supported on a finite set $S$ of edges.
Assume that $\lambda$ belongs to the interior of a spectral band
of $H$ and is not a Dirichlet eigenvalue on any of the edges, the
corresponding Floquet surface is irreducible, and $\lambda$ is an
embedded eigenvalue for $H+w$. Then the corresponding
eigenfunction $f\in L_2(\GG)$ of $H+w$ is compactly supported and
moreover,
$$
r(supp\,f) <
r(\widetilde{S})+r(\widetilde{W_1})(2|\widetilde{W_1}|+1).
$$
\end{theorem}

{\bf Proof of Theorem \ref{T:main_quantum1}} (and therefore also
of Theorem \ref{T:main_quantum2} and Theorem \ref{T:main_quantum})
is based upon its reduction to the discrete version Theorem
\ref{T:main}.

Since we are guaranteed that a neighborhood of $\lambda$ is free
of Dirichlet spectra of individual edges, one can use the standard
procedure of reducing the spectral problems for $H$ and for $H+w$
for the quantum graph to the one for a combinatorial one (e.g.,
\cite{Al,Kuch_graphs1,Kuch_graphs2}). Namely, assume that one
solves the following problem on the graph:
\begin{equation}\label{E:sp_problem}
    \begin{cases}
    -\frac{d^2f}{dx^2}+V(x)f=\lambda f \mbox{ on each edge}\\
    $f$ \mbox{ is continuous and } \sum\frac{df}{dx_e}=0 \mbox{ at each
    vertex}.
    \end{cases}
\end{equation}
Since $\lambda$ is not in the Dirichlet spectrum on each edge, one
can solve the equation on each edge, assuming the knowledge of
vertex values of the function, and then express the Neumann data
at the vertices as a linear combination of vertex values. Thus,
the only condition that still remains to be satisfied is that the
outgoing derivatives at each vertex add up to zero. This clearly
will take form of an equation
\begin{equation}\label{E:vertex_problem}
    \sum\limits_{v\sim u} a_{u,v}(\lambda)f(v)=0
\end{equation}
satisfied at each vertex $u$ of the graph. One sees that this can
be written as a second order difference equation $A(\lambda)f=0$
on the combinatorial graph. Notice that $supp_W(A(\lambda))\subset
\widetilde{W}$. Analogously, the perturbed equation can be
rewritten as $A_1(\lambda)f=0$. This leads to the two
combinatorial counterparts of our periodic and perturbed spectral
problems:
\begin{equation}\label{E:vertex_sp_problem}
A(\lambda)f=0, A_1(\lambda)f=0.
\end{equation}
In order to prove the theorem, we will need some simple auxiliary
statements collected in the following
\begin{lemma}\label{L:transfer}
\begin{enumerate}
\item If a function $f$ on the quantum graph satisfies $Hf=\lambda
f$ (resp. $(H+w)f=\lambda f$), then its vertex values satisfy the
difference equations $A(\lambda)f=0$ (resp. $A_1(\lambda)f=0$).
Conversely, if a vector $f$ of vertex values satisfies
$A(\lambda)f=0$ (resp. $A_1(\lambda)f=0$), it can be uniquely
extended to a solution of $Hf=\lambda f$ (resp. $(H+w)f=\lambda
f$).

\item If the values of such a solution $f$ at both vertices of an
edge are equal to zero, then $f$ is zero on this edge. In
particular, $f$ is compactly supported if and only if its vertex
values are compactly supported, and both supports are of
equivalent sizes (i.e., their radii are the same).

\item The operator $A(\lambda)$ is periodic.

\item The difference operator $B=A_1(\lambda)-A(\lambda)$ is
supported only on the vertices that are incident to the edges
where $w$ has a non-empty support. In particular, $A_1(\lambda)$
is a compactly supported perturbation of $A(\lambda)$ with the
size of the support of the perturbation controlled by the size of
the support of $w$.

\item The Floquet surfaces satisfy the following relation:
\begin{equation}\label{E:Floquet_same}
    \Phi_{H,\lambda}=\Phi_{A(\lambda),0}.
\end{equation}
\end{enumerate}
\end{lemma}
{\bf The proof of the lemma} is rather straightforward. Indeed,
the way the operators $A$ and $A_1$ are defined implies the direct
part of the first claim of the lemma. The converse part is also
simple. Indeed, if a vector $f$ of vertex values satisfies
$A(\lambda)f=0$, let us solve the equation $Hu=\lambda u$ on each
edge taking $f$ as Dirichlet boundary values (this is possible due
to our avoidance of Dirichlet spectra). The resulting function
satisfies the equations on the edges and continuity condition by
construction. The Neumann condition at the vertices follows now
from $A(\lambda)f=0$.

The second claim of the lemma follows from the same avoidance of
the Dirichlet spectra.

The third statement follows from periodicity of $H$.

The fourth statement is straightforward from the definitions of
$A(\lambda)$ and $A_1(\lambda)$.

Let us prove the important (albeit still simple) last statement.
If $z\in \Phi_{H,\lambda}$, this means that there exists a
non-zero function $f$ satisfying the equation $Hf=\lambda f$ and
such that $f(gx)=z^g f(x)$ for any $g\in \ZZ^n$. This cyclic
relation in particularly holds at the vertices, which implies that
$z\in \Phi_{A(\lambda),0}$. Conversely, if $z\in
\Phi_{A(\lambda),0}$, then there is a cyclic vertex function $u$
with the Floquet multiplier $z$ such that $A(\lambda)u=0$. Let us
use it as Dirichlet data on each edge to solve
$-\frac{d^2f}{dx^2}+V(x)f=\lambda f$ on each edge. The first claim
of the lemma guarantees that we get a solution $f$ of $Hf=\lambda
f$. We claim that $f$ is cyclic with the Floquet multiplier $z$.
Indeed, for any $g\in \ZZ^n$ the functions $f(gx)$ and $z^gf(x)$
satisfy the same equation $-\frac{d^2f}{dx^2}+V(x)f=\lambda f$ on
each edge and have the same Dirichlet data. Since we avoided
Dirichlet spectrum, we conclude that $f(gx)=z^gf(x)$. This proves
the Lemma. \qed

We can now finish the proof of the Theorem \ref{T:main_quantum}.
Indeed, the previous Lemma guarantees that switching from the
differential periodic and perturbed problems $Hf=\lambda f$ and
$(H+w)f=\lambda f$ to the combinatorial problems $A(\lambda)f=0$
and $A_1(\lambda)f=0$, one lands in the conditions of Theorem
\ref{T:main}. Now the same lemma implies that the conclusion of
the Theorem \ref{T:main} about the vertex values implies the
conclusion of Theorem \ref{T:main_quantum} about the whole
function $f$. \qed

\section{Remarks and acknowledgments}

\begin{itemize}

\item The notion of the ``radius'' $r(S)$ of a finite set in $\GG$
depends on the choice of a fundamental domain $W$. Indeed,
choosing $W$ further away from $S$ increases $r(S)$. Thus, the
optimal way to use the estimates of the main theorems is to choose
a fundamental domain $W$ in such a way that $r(S)$ is the smallest
possible for a given support $S$ of the perturbation. This would
lead to the best localization estimate for the embedded
eigenfunctions.

\item As it has already been mentioned in the previous section,
the Neumann-Kirchhoff conditions (\ref{E:conditions}) are chosen
for simplicity of exposition sake only. Since the results
concerning combinatorial graphs are obtained under very general
conditions on the periodic operator $A$, the statement of Theorem
\ref{T:main_quantum} and its proof carry through if one uses the
general self-adjoint vertex conditions (described for instance in
\cite{Harmer,KoS1,Kuch_graphs1}). One might have to deal with
matrix difference operators $A$, which causes no problem. However,
the specific estimates of the constant $C$ of Theorem
\ref{T:main_quantum} given in Theorems \ref{T:main_quantum1} and
\ref{T:main_quantum2} will have to change depending on the vertex
conditions.

\item A deficiency of the results of this paper (as well as of the
results of \cite{KuchVain1,KuchVain2}) is that our technique does
not let us treat the case of eigenvalues embedded at spectral
edges.

\item It is clear from both this paper and
\cite{KuchVain1,KuchVain2} that question of irreducibility
 of the Floquet surface (equivalently, of the Fermi surface, modulo
 natural periodicity) is intimately related to the problem of existence
 and behavior of embedded eigenvalues and corresponding eigenfunctions.
 This does not look like an artifact of the techniques used. It is clear
 that not arbitrary periodic
 difference operator satisfies this condition (even higher order periodic
 elliptic differential operators do not \cite{Kuch_book}). However, it has
 been proven for discrete Schr\"{o}dinger operator $-\Delta +V(x)$ on $\ZZ^2$
 with potential periodic with respect to a sublattice, that
 irreducibility holds except possibly for finitely many values of
 $\lambda$ \cite{Gieseker}. It was also conjectured in the book
 \cite{Gieseker} that irreducibility always holds for this
 operator (similar statement is probably correct in any
 dimension).

 Irreducibility is also known for such operators (both discrete and continuous)
 with separable potentials \cite{BKT,Gieseker} (see also \cite{KuchVain1}). It is even
 sufficient to have in dimension three the potential to separate
 as $V_1(x_1)+V_2(x_2,x_3)$ \cite{BKT,Gieseker,KuchVain1}. This
 follows from the results on irreducibility of Bloch variety in
 dimension two \cite{Knorrer}.

 An advantage of dealing with a difference operator is a
 possibility of sometimes explicitly computing the determinant $\Delta(z)$
 and thus checking its irreducibility.

 In fact, examination of the proofs of this text, as well as of \cite{KuchVain1,KuchVain2}
 shows that we do not need complete irreducibility. What truly required, is that every
 irreducible component of the Floquet variety intersects the torus $\TT^n$ over an $n-1$
 dimensional set. However, it is not clear how to control this property, and thus it is doubtful
 that such weaker condition will work better than the full irreducibility.

\item As it has been mentioned already, pathologies like pure
point spectrum of periodic operators and embedded eigenvalues
might and do sometimes appear in discrete or quantum graph
situation. However, they do not necessarily have to. Indeed, it is
known \cite{KuchDifference} that the discrete Schr\"{o}dinger
operator $-\Delta +V(x)$ on the lattice $\ZZ^n$ with a potential
periodic with respect to a sublattice has absolutely continuous
spectrum. This can be proven by the standard L.~Thomas' argument
\cite{Thomas}. Similarly, there are some cases when one can prove
that embedded eigenvalues do not arise from local perturbations of
periodic discrete operators. Assume for instance that a difference
operator $P$ on the integer lattice $\ZZ^n$ (the operator could in
particular be our perturbed operator $A+B$) has the following
property: there exists an oriented hyperplane $L$ in $\RR^n$ such
that for any point $y\in\GG$ there exists a point $x\in\GG$ such
that $supp_x(P)$ contains the point $y$ and lies completely on the
``positive'' side of the parallel shift $L_y$ to the point $y$,
with the only intersection with $L_y$ at $y$. Then the equation
$Pf=0$ has no compactly supported solutions. Indeed, if there were
such a solution $f$, consider a support hyperplane $L_y$ to
$supp(f)$ such that the whole $supp(f)$ is on the negative side of
$L_y$ and $y\in supp(f)$. Consider the point $x$ that serves $y$
as described above. Then the equality $(Pf)(x)=0$ clearly implies
that $f(y)=0$, which is a contradiction.

This in particular proves the quoted above fact about absence of
point spectrum for periodic Schr\"{o}dinger operators on integer
lattices.

\item It would be interesting to understand how much the
assumption of commutativity of the group of periods influences the
validity of the results of this paper. We do not know the answer
to this question, but one probably should not expect to be able to
go beyond the class of groups of polynomial growth (and hence,
according to M.~Gromov's result \cite{Gromov}, virtually nilpotent
ones). Indeed, the results already quoted about the unusual
spectral behavior of the lamplighter group (which is of an
intermediate growth) \cite{GrigZuk,dicks} show that one might
expect surprises there.

\item The approach used in this work has been previously used by
the authors in different circumstances in
\cite{KuchVain1,KuchVain2} (see also \cite{Shaban}). Its idea
originates from the paper \cite{Vainberg} of the second author.

\end{itemize}

This research was partly sponsored by the NSF through the Grants
DMS-0406022 (the first author) and DMS-0405927 (the second
author). The authors thank the NSF for this support. The content
of this paper does not necessarily reflect the position or the
policy of the federal government, and no official endorsement
should be inferred.

\end{document}